\documentclass[aps,prl,showpacs,twocolumn]{revtex4}

\newcommand{\be}{\begin{eqnarray}}
\newcommand{\ee}{\end{eqnarray}}
\newcommand{\no}{\nonumber}
\newcommand{\Str}{{\rm Str}\,}

\begin{document}
\title{Bosonization for disordered and chaotic systems}
\author{K. B. Efetov${}^{1,2}$, G. Schwiete${}^{1}$, and K. Takahashi${}^{1}$}
\affiliation{${}^1$ Theoretische Physik III, 
 Ruhr-Universit\"at Bochum, 44780 Bochum, Germany \\
 ${}^2$ L. D. Landau Institute for Theoretical Physics, 117940 Moscow, Russia}
\date{\today}
\begin{abstract}
 Using a supersymmetry formalism, we reduce exactly the problem of electron motion 
 in an external potential to a new supermatrix model valid at all distances. 
 All approximate nonlinear $\sigma$ models obtained previously for disordered systems 
 can be derived from our exact model using a coarse-graining procedure. 
 As an example, we consider a model for a smooth disorder 
 and demonstrate that using our approach does not lead to a ``mode-locking'' problem. 
 As a new application, we consider scattering on strong impurities for which 
 the Born approximation cannot be used. 
 Our method provides a new calculational scheme for disordered and chaotic systems.
\end{abstract}
\maketitle


 The supersymmetry method has proven to be a useful method  
 for solving different problems for disordered systems \cite{Efetov}.  
 Within this technique, for the problem of weakly disordered conductors,  
 one derives a supermatrix nonlinear $\sigma$ model  
 which is valid for distances exceeding the transport mean free path $l_{\rm tr}$ 
 and is applicable for a description of the diffusive motion of electrons.
 The main object in this approach is an 8$\times$8 supermatrix field $Q({\bf r})$  
 that depends on the coordinate ${\bf r}$.  
 The free energy functional for space variations of $Q({\bf r})$ has the form 
\be
 F[Q] = \frac{\pi\nu}{8}\int\Str \left[ 
 D(\nabla Q)^2+2i(\omega+i\delta)\Lambda Q\right] d{\bf r}, \label{dif}
\ee
 where $D$ is the classical diffusion coefficient, 
 $\nu$ is the density of states at the Fermi surface, and $\omega$ is the frequency.  
 The matrix $\Lambda={\rm diag}(1,-1)$ distinguishes between matrix elements 
 corresponding to advanced and retarded Green functions 
 and $\Str$ means supertrace.  
 The supermatrix $Q({\bf r})$ satisfies the relation $Q^2({\bf r})=1$.

 Recently, a lot of attention has been paid to the study of ballistic electron motion 
 and related problems of quantum chaos (for a review see, e.g., Ref.\cite{LKK}). 
 Equation (\ref{dif}) is not applicable in this case and 
 one needs a generalization of the nonlinear $\sigma$ model. 
 It is expected that the proper model can be applied to clean systems 
 which are chaotic in the classical limit. 
 Treating disordered and chaotic systems in the same manner is of great advantage 
 since they share common properties such as the universality of the level statistics, 
 although the origin of randomness is totally different.

 In these problems, the traditional derivation of the nonlinear $\sigma$ model 
 based on the disorder averaging is of no use and one needs more general derivation. 
 Several attempts to generalize the $\sigma$ model and include shorter distances 
 have been undertaken in the past several years \cite{MK,AASA,Zirn,TE,EK}. 
 Although this study has lead to considerable progress in understanding 
 the ballistic motion, the goal of the complete description has not been achieved. 
 We mention that the most general quasiclassical form of the ballistic $\sigma$ model 
 has been written for a smooth disorder potential in a recent work \cite{EK}. 
 It is valid for distances down to the wavelength $\lambda_{\rm F}$. 
 However, it seems that even shorter distances are also important 
 because in the absence of a disorder, the approach of Ref.\cite{EK} 
 (as well as all others) does not allow one to derive a so-called ``regularizer'' 
 of Refs.\cite{AL,AOS} that is believed to originate from the quantum diffraction 
 and is absolutely necessary to obtain the Ehrenfest time.

 In this Letter, we suggest a scheme of derivation of the $\sigma$ model that is 
 completely different from all those of Refs.\cite{sigma,Efetov,MK,AASA,AOS,Zirn,TE,GM,EK}. 
 In contrast to the previous derivations, we are able to reduce 
 the initial electron problem to a model containing an integral over 
 supermatrices $Q({\bf r},{\bf p})$ exactly. 
 This model is a generalization of the $\sigma$ model to all distances 
 and is written for any given potential. 
 It can be simplified by integrating out shorter distances and 
 this is how semiclassical ballistic $\sigma$ models and 
 the diffusive model (\ref{dif}) can be derived.

 The method we use is some kind of the bosonization that maps the original
 fermionic system onto a bosonic one, which has been used for a wide variety
 of physical systems \cite{boson1}. 
 Our approach is also known as the geometrical quantization or the coherent-state method 
 and is very close to that in Ref.\cite{boson2} 
 (a similar trick has been used in a proof of the universality 
 for random matrices \cite{HW}). 
 At the same time, the supersymmetry allows us to do considerably more and 
 obtain a comparatively simple explicit form for the bosonized action.


 Using $8$-component supervectors $\psi({\bf r})$  
 we write the generating function $Z[J]$ for the electron systems as \cite{Efetov} 
 $Z[J] = \int\exp(-{\cal L})\mathcal{D}(\psi ,\bar{\psi})$ with 
\be
 {\cal L} &=& -i\int\bar{\psi}({\bf r}){\cal H}_{J}\left({\bf r},{\bf r}^{\prime}\right)
 \psi({\bf r}^{\prime}) d{\bf r}d{\bf r}^{\prime}, \\
 {\cal H}_{J} &=& \delta\left({\bf r}-{\bf r}^{\prime}\right) 
 \left(\hat{H}-\varepsilon+\frac{\omega+i\delta}{2}\Lambda\right) 
 -J\left({\bf r},{\bf r}^{\prime}\right),
\ee
 where $\hat{H}$ is the Hamiltonian of the electron system and 
 $J({\bf r},{\bf r}^{\prime})$ is a source field.  
 The variables $\varepsilon$ and $\omega$ stand for the electron energy  
 and the frequency of the external field, respectively.  
 We write the Hamiltonian as $\hat{H}=\hat{\bf p}^2/2m-\varepsilon_{\rm F}+U({\bf r})$,  
 where $\varepsilon_{\rm F}$ is the Fermi energy and $U({\bf r})$ is a potential 
 that may include a regular part as well as an impurity potential.
 An extension to more complicated systems including magnetic or  
 spin-orbit interactions is straightforward.

 Now we want to express the original generating function $Z[J]$ in a ``bosonized'' form. 
 We start our derivation with the identity 
\be
 1=\int e^{-i\int\Str\left[(1/2)Q({\bf r},{\bf r}^{\prime})
 -\psi({\bf r})\bar{\psi}({\bf r}^{\prime})\right] 
 A({\bf r}^{\prime},{\bf r})d{\bf r}d{\bf r}^{\prime}} {\cal D}Q{\cal D}A,
\ee
 where $Q$ and $A$ are arbitrary supermatrices with the same symmetry 
 as the product $\psi({\bf r})\bar{\psi}({\bf r}^{\prime})$. 
 This identity is evident because the integral over the variable $A$ 
 gives the delta function $\delta(Q/2-\psi\bar{\psi})$. 
 The factor $1/2$ is written to keep along the notations of Ref.\cite{Efetov}. 
 Substituting this expression into $Z[J]$ we integrate over the supervectors $\psi({\bf r})$
 and shift the variable $A$ as $A\rightarrow A_{\rm sp}+A$, 
 where $A_{\rm sp}={\cal H}_{J}-iQ^{-1}$. 
 Then we obtain 
\be
 Z[J] &=& \int\exp\left\{-\frac{i}{2}\int\Str[{\cal H}_{J}({\bf x})\ast 
 Q({\bf x})]d{\bf x}\right\}{\cal B}[Q]{\cal D}Q, \no\\ 
\ee
\begin{widetext}
\be
 {\cal B}[Q] = \int\exp\biggl(\frac{1}{2}\int\Str 
 \{\ln [iQ^{-1}({\bf x})-A({\bf x})] 
 -iA({\bf x})\ast Q({\bf x})\}d{\bf x}\biggr){\cal D}A.
\ee
\end{widetext}
 All functions in the above equations are written using the Wigner representation 
 $A({\bf R},{\bf p})=\int d{\bf \rho}A({\bf r},{\bf r}^{\prime})
 \mbox{e}^{-i{\bf p\rho}}$, where ${\bf R}=({\bf r}+{\bf r}^{\prime})/2$ and 
 ${\bf \rho}={\bf r}-{\bf r}^{\prime}$, and we use the notation 
 ${\bf x}=({\bf r},{\bf p})$ for a point in the phase space. 
 The products of two supermatrices $A({\bf x})\ast B({\bf x})$ 
 are defined by the Moyal formula 
\be
 A({\bf x})\ast B({\bf x}) =  A({\bf x})\mbox{e}^{(i/2)(
 \stackrel{\leftarrow}{\nabla}_{\bf r}\stackrel{\rightarrow}{\nabla}_{\bf p}
 -\stackrel{\rightarrow}{\nabla}_{\bf r}\stackrel{\leftarrow}{\nabla}_{\bf p}
 )}B({\bf x}).
\ee
 Calculation using the Wigner representation and the star product $\ast$ is known as 
 the Weyl symbol calculus \cite{McD} and is convenient for quasiclassical expansions.

 The generating function $Z[J]$ is expressed in terms of 
 the integral over the variables $Q$ and $A$. 
 Remarkably, due to the supersymmetry 
 one can calculate the integral over $A$ exactly. 
 This can be done writing the supermatrix $Q({\bf x})$ as 
 $Q({\bf x})=T({\bf x})\ast q({\bf x})\ast\bar{T}({\bf x})$, 
 where $q({\bf x})$ is a diagonal matrix and $T({\bf x})\ast\bar{T}({\bf x})=1$. 
 Changing the variables $A({\bf x})\rightarrow T({\bf x})\ast A({\bf x})\ast \bar{T}({\bf x})$ 
 we express the integral ${\cal B}$ as a function of the variable $q({\bf x})$ only. 
 We can proceed further making one more change of the variables of the integration 
 $A({\bf x})\rightarrow q^{-1/2}({\bf x})\ast A({\bf x})\ast q^{-1/2}({\bf x})$. 
 The Jacobian arising when changing the variables is unity due to the supersymmetry. 
 Then we find the new variable $A$ does not couple to $Q$. 
 Because of the supersymmetry the final integral over $A$ 
 is exactly equal to unity and
 we obtain the generating function $Z[J]=\int\exp(-F[Q]){\cal D}Q$, where 
\be
 F[Q] = \frac{i}{2}\int\Str\left[ 
 {\cal H}_J({\bf x})\ast Q({\bf x})-i\ln Q({\bf x})\right] d{\bf x}.  \label{F}
\ee

 This is the main result of the present work.  
 We see that the initial electron problem which is written in terms of  
 the supervector $\psi({\bf r})$ is reduced exactly to a functional integral over  
 the supermatrices $Q({\bf x})$ depending on the coordinates ${\bf x}$ in the phase space.  
 Since $\psi({\bf r})$ are supervectors and not fermionic vectors,  
 we may call the procedure leading to Eq.(\ref{F}) ``superbosonization''.

 We emphasize that the supersymmetry is used in this formulation for writing 
 the generating function in terms of an integral over matrices but not for averaging. 
 No averaging over impurities has been performed in the derivation of  
 the functional $F[Q]$ and the derivation itself is just  
 a sequence of identical transformations.  
 This contrasts the traditional supersymmetry method \cite{Efetov}  
 where the supersymmetry was used for averaging over disorder  
 in the beginning of the calculations.  
 The variables $Q({\bf x})$ in $F$ are different from  
 the variables of integration used in other related works  
 and the representation (\ref{F}), being as exact as the original representation,  
 is simply more convenient for making approximations.  
 Its validity is by no means related to a form of the disorder distribution.


 Now we want to show how the formalism works in different situations.  
 First, let us consider a system with a smooth potential such that  
 one may expect a quasiclassical behavior.  
 We demonstrate how one can come to the model of Ref.\cite{EK} 
 starting from Eq.(\ref{F}).  
 We do not discuss here clean billiards and how one should average over the spectrum.  
 This is an important and complicated problem but it is beyond the scope of the present Letter.  
 Instead, we consider, as in Refs.\cite{AL,EK}, an infinite system with a smooth disorder.

 The functional $F[Q]$ is quite complicated and 
 we find first its minimum in $Q({\bf x})$ at $J=0$. 
 The minimum is achieved at $q_0({\bf x})=i{\cal H}_{J=0}^{-1}=g({\bf x})$, 
 where $g({\bf x})$ is the one particle Green function of the Hamiltonian 
 without sources (multiplied by $i$). 
 If the potential $U({\bf r})$ in the Hamiltonian is smooth, 
 the dependence of $g({\bf r},{\bf r}^{\prime})$ on $({\bf r}+{\bf r}^{\prime})/2$ 
 is weaker than on ${\bf r}-{\bf r}^{\prime}$ and we can write with good accuracy 
\be
 \frac{1}{\pi}\int g({\bf x})d\xi \simeq 
 \frac{i}{\pi}\int\frac{d\xi}{{\bf\xi}({\bf p})+i\delta\Lambda} \simeq \Lambda, \label{qclg}
\ee
 where $\xi({\bf p})={\bf p}^2/2m-\varepsilon_{\rm F}$.
 Writing this equation we assumed that the spectrum is continuous, 
 which is definitely true for an infinite metallic sample. 
 In a finite system, the use of $g({\bf x})$ in the form of Eq.(\ref{qclg}) 
 may be justified after a certain averaging over impurities.

 In order to compute the contribution of fluctuations near the saddle point
 we write $Q({\bf x})$ in Eq.(\ref{F}) as 
 $Q({\bf x})=T({\bf x})\ast q({\bf x})\ast \bar{T}({\bf x})$,
 where $q({\bf x})$ is a block-diagonal supermatrix 
 (commuting with $\Lambda$) and $T\ast\bar{T}=1$. 
 If $J=0$, fluctuations of the supermatrix $q({\bf x})$ near $q_0$ 
 do not give any contribution and its average is equal to 
 the exact one particle Green function $g({\bf x})$. 
 This is because fluctuations of $T$ do not contribute to 
 the one particle Green functions \cite{Efetov} and 
 one can simply put $T=1$ calculating this quantity. 
 Then we immediately come to the conclusion that 
 $\left<q({\bf x})\right>=g({\bf x})$, which means that the integration 
 over the ``massive'' modes (fluctuations of $q$) does not give any correction.

 In contrast, fluctuations of $T({\bf x})$ are important  
 and the main contribution comes from matrices slowly varying in the phase space.  
 We perform first the integration over supermatrices $q$  
 by making a cumulant expansion in gradients of $T$.  
 The main contribution comes from the first cumulant.  
 Putting $q({\bf x})=g({\bf x})$ in Eq.(\ref{F}),  
 we can easily integrate over $\left|{\bf p}\right|$  
 because only $g({\bf x})$ should be integrated,  
 whereas the momentum ${\bf p}$ in the supermatrices $T({\bf x})$  
 should be taken at the Fermi surface, ${\bf p}=p_{\rm F}{\bf n}$, ${\bf n}^2=1$.  
 Expanding the star products in gradients,  
 which corresponds to the quasiclassical approximation, we reduce $F$ to the form 
\be
 F &=& \frac{\pi\nu}{2}\int\Str\biggl\{
 \Lambda\bar{T}_{\bf n}({\bf r})
 \left[v_{\rm F}{\bf n}\nabla_{\bf r}
 -p_{\rm F}^{-1}\nabla_{\bf r}U({\bf r})\nabla_{\bf n}\right]
 T_{\bf n}({\bf r})
 \no\\
 & & 
 +i\left[\frac{\omega+i\delta}{2}\Lambda-J_{\bf n}({\bf r})\right] 
 Q_{\bf n}({\bf r}) 
 \biggr\}d{\bf r}d{\bf n}, \label{Fqc}
\ee
 where $Q_{\bf n}({\bf r})=T_{\bf n}({\bf r})\Lambda\bar{T}_{\bf n}({\bf r})$.  
 This functional 
 agrees with the corresponding functional  
 derived from an equation for quasiclassical Green functions \cite{EK}.  
 Equation (\ref{Fqc}) is applicable for all distances 
 exceeding the wavelength $\lambda_{\rm F}$.  
 One can find a detailed investigation of properties of this functional and  
 its reduction to Eq.(\ref{dif}) in Ref.\cite{EK}.

 It is clear from the derivation that the momenta ${\bf p}$ are pinned in 
 the supermatrices $T({\bf x})$ to the Fermi surface due to the singular form  
 of the supermatrix $q({\bf x})=g({\bf x})$ near the Fermi surface.  
 In the traditional approach based on 
 the Hubbard-Stratonovich or color-flavor transformation,  
 the matrix $q({\bf x})$ 
 is smooth at the Fermi surface and nothing suppresses fluctuations 
 perpendicular to the energy shell (the ``mode-locking'' problem).  
 We see that the mode-locking problem is related merely to an inconvenient choice  
 of the variables of the integration in the traditional approach  
 and this problem does not exist here  
 (it does not exist in the diagrammatic approach \cite{AL} and  
 the approaches based on the quasiclassical equations \cite{MK,EK}).

 In the absence of the random potential $U({\bf r})$ one has only terms 
 linear in gradients and no regularizer such as that suggested 
 in Refs.\cite{AASA,AOS,AL} appears in Eq.(\ref{Fqc}),
 which would mean that classical orbits do not mix.
 However, we are sure that a more careful consideration of quantum effects
 leading to the mixing of the classical motion
 is possible within the model (\ref{F}).


 Now let us consider a disordered metal with 
 a small concentration of strong short-ranged impurities, 
 such that the self-consistent Born approximation(SCBA) cannot be used. 
 This problem cannot be considered by the traditional $\sigma$ model scheme of 
 Ref.\cite{Efetov} because the saddle point approximation used in this approach 
 is equivalent to the SCBA.
 Moreover, we are not aware of any diagrammatic calculation of 
 the weak localization correction for this model, 
 although the effect of strong impurities was considered 
 in different situations \cite{LO,Altland}.

 Our derivation of the reduced $\sigma$ model consists of three steps: 
 the coarse-graining procedure, the quasiclassical approximation, 
 and the impurity averaging. 
 First, we integrate out short distances of the order of $\lambda_{\rm F}$ and 
 reduce the initial model to an effective one describing fluctuations at longer distances. 
 After this coarse-graining procedure, we can use the quasiclassical approximation. 
 At the same time, the impurities are self-averaged over the integrated length scale. 
 It is important to notice that this method of the averaging is different from 
 the traditional method where the impurity is averaged 
 by considering the ensemble of systems. 
 Since all of the variables are slowly varying in space,
 we can simply average the terms of the coarse-grained functional 
 by changing the observation point, 
 which is equivalent to the averaging over the positions of the impurities.

 In order to perform the coarse-graining procedure, 
 we rewrite the supermatrix $Q$ as 
 $Q({\bf x})=V({\bf x})\ast\tilde{Q}({\bf x})\ast \bar{V}({\bf x})$, 
 where $V=\bar{V}^{-1}$ describes slow modes and $\tilde{Q}$ stands for fast modes. 
 Substituting this representation into Eq.(\ref{F}) 
 we write $\tilde{Q}({\bf x})=g({\bf x})+\delta Q({\bf x})$
 and expand $F$ in $\delta Q$ up to the second order. 
 Taking into account fast modes [fluctuations of $\tilde{Q}({\bf x)}$]  
 is necessary to obtain the correct form of the collision term. 
 This is done by performing Gaussian integrals over $\delta Q$. 
 Alternatively, we could use the contraction rules for $Q$ 
 which can be obtained by comparing $Z[J]$ written in terms of $\psi$ 
 and $Z[J]$ of $Q$.
 This allows us to take into account higher order effects in a systematic way. 
 As a result, we obtain 
\be
 F &=& \frac{i\omega}{4}\int\Str\Lambda Q d{\bf x}
 +\frac{i}{2}\int\Str H\ast Q d{\bf x}  \no\\
 & & -\frac{1}{4}\int\Str\left(g\ast\bar{V}\ast
 \left[H\stackrel{\ast}{,}V\right]\right)^2 d{\bf x},  \label{Fcg}
\ee
 where the $Q$ matrix has been reduced to 
 $Q({\bf x})=V({\bf x})\ast g({\bf x})\ast \bar{V}({\bf x})$. 
 We neglected the source term for simplicity.

 Since the variables $V({\bf x})$ are slow we can use the quasiclassical approximation. 
 This means that we introduce the quasiclassical Green function 
 $g_{\bf n}({\bf r})=\pi^{-1}\int g({\bf x})d\xi$ 
 as Eq.(\ref{qclg}) and expand the star product. 
 The Green function $g({\bf x})$ still includes the impurity potential and 
 we introduce the function $X({\bf x})$ as
 $g({\bf x})=g_0({\bf p})\ast [1+iX({\bf x})\ast g_0({\bf p})]$,
 where $g_0({\bf p})$ is the one particle Green function 
 in the absence of impurities.
 Writing the impurity potential as a sum over the positions ${\bf r}_a$
 of the impurities $U({\bf r})=\sum_a u({\bf r}-{\bf r}_a)$,
 where $u({\bf r})$ is the potential of a single impurity, we average over ${\bf r}_a$. 
 A standard consideration (see, e.g., \cite{AGD}) leads for a small impurity concentration 
 $n_{\rm i}$ to the following result in the momentum representation 
\be
 \left<X({\bf p},{\bf p}^{\prime})\right>_{\rm imp} = 
 \frac{n_{\rm i}f({\bf p},{\bf p})}{1-ig_0({\bf p})n_{\rm i}f({\bf p},{\bf p})}
 \delta ({\bf p}-{\bf p}^{\prime}),
\ee
 where $f$ is the exact amplitude of scattering on a single impurity. 
 Then we find for the Green function 
 $\left<g({\bf p}+{\bf q}/2,{\bf p}-{\bf q}/2)\right>_{\rm imp} = 
 i\delta({\bf q})/(\xi({\bf p})+i\Lambda/2\tau)$,
 and for the quasiclassical Green function, $g_{\bf n}({\bf r})=\Lambda$. 
 We introduced the scattering time $\tau$ using the optical theorem 
 ${\rm Im}f({\bf n},{\bf n})=\pi\nu\Lambda\int d{\bf n}^{\prime}
 |f({\bf n},{\bf n}^{\prime})|^{2}=(2n_{\rm i}\tau)^{-1}\Lambda$. 
 Other contributions such as the average of a product of $X$ can be considered 
 in a similar way and are expressed in terms of the scattering amplitude $f$. 

 We can write the coarse-grained functional (\ref{Fcg}) using the scattering amplitude 
 $f({\bf p},{\bf p}^{\prime})$ and the free particle Green function $g_0({\bf p})$ 
 instead of using the impurity potential $U({\bf r})$ 
 and the full Green function $g({\bf x})$. 
 Introducing the quasiclassical Green function, 
 the momentum ${\bf p}$ is restricted to the Fermi surface. 
 We write the supermatrix $V$ in the form 
 $V_{{\bf n}}({\bf r})=V({\bf r})[1-iP_{\bf n}({\bf r})-(1/2)P_{\bf n}^2({\bf r})+\cdots]$, 
 $P\Lambda+\Lambda P=0$ and expand Eq.(\ref{Fcg}) in $P$.
 We obtain 
\be
 F &=& \frac{\pi\nu}{2}\int\Str\biggl\{
 \Lambda\bar{V}_{\bf n}({\bf r})
 v_{\rm F}{\bf n}\nabla_{\bf r}V_{\bf n}({\bf r})
 +\frac{i\omega}{2}\Lambda Q_{\bf n}({\bf r}) \no\\
 & & +w({\bf n},{\bf n}^{\prime})
 \left[P_{\bf n}({\bf r})-P_{{\bf n}^{\prime}}({\bf r})\right]^{2}\biggr\}
 d{\bf r}d{\bf n}d{\bf n}^{\prime},   \label{Fb}
\ee
 where $w({\bf n},{\bf n}^{\prime})=2\pi\nu\left|f({\bf n},{\bf n}^{\prime})\right|^{2}$ 
 is the probability of scattering from the state with ${\bf n}$ to the state 
 with ${\bf n}^{\prime}$, and 
 $Q_{\bf n}({\bf r})=V_{\bf n}({\bf r})\Lambda\bar{V}_{\bf n}({\bf r})$. 
 In Eq.(\ref{Fb}), we keep in the collision (last) term second powers of $P$ only. 
 When averaging over the impurities, it is difficult to keep all powers of $P$ and 
 this is the reason why we do not write higher order terms. 
 Equation (\ref{Fb}) is not sufficient 
 for the complete description of the ballistic regime 
 where the collision term is, generally speaking, a complicated function of $P$. 
 However, as long as we are interested in the diffusive regime, 
 Eq.(\ref{Fb}) is enough and we can come to Eq.(\ref{dif}) 
 by a Gaussian integration over the supermatrices $P$. 
 The diffusion coefficient $D$ in Eq.(\ref{dif}) is given by 
 $D=v_{\rm F}^2\tau_{\rm tr}/d$ ($d$ is the dimensionality), 
 where the transport time $\tau_{\rm tr}$ is defined as 
 $\tau_{\rm tr}^{-1}=n_{\rm i}\int w({\bf n},{\bf n}^{\prime})
 (1-{\bf n}{\bf n}^{\prime})d{\bf n}^{\prime}$, 
 and $Q\left({\bf r}\right)=V({\bf r})\Lambda\bar{V}({\bf r})$. 


 In conclusion, we developed a new approach to disordered and chaotic systems. 
 The most general supermatrix model obtained in Eq.(\ref{F}) is exact 
 and valid at all distances. 
 It can be simplified by using a coarse-graining procedure 
 for both smooth disorder and strong impurities, 
 which leads to ballistic and diffusive nonlinear $\sigma$ models. 
 We believe that our method is a proper route to 
 a more complete description of the quantum chaos.

 We acknowledge the support of the SFB/Transregio 12 and GRK 384.  
 We are grateful to V.R. Kogan, A.I. Larkin and P.B. Wiegmann for useful discussions.



\end{document}